\documentstyle[epsfig,preprint,aps]{revtex}
\begin{document}
\tighten
\draft
\title{Meson mixing amplitudes in asymmetric nuclear matter}
%
\author{Yoshiharu Mori$^a$\footnote{E-mail: mori@nucl.phys.tohoku.ac.jp}
and Koichi Saito$^b$\footnote{ksaito@tohoku-pharm.ac.jp}
}
\address{$^a$ Department of Physics, Tohoku University, Sendai 980-0845,
Japan \\
$^b$ Tohoku College of Pharmacy, Sendai 981-8558, Japan }
\maketitle
\begin{abstract}
Using a purely hadronic model, we study the charge-symmetry-breaking 
$\rho$-$\omega$, $\sigma$-$\delta$, $\sigma$-$\rho$ and
$\delta$-$\omega$ mixing amplitudes 
in isospin asymmetric nuclear matter.  The basic assumption 
of the model is that the mixing amplitude is generated by nucleon and
anti-nucleon 
loops and hence driven entirely by the difference between proton and
neutron Fermi momenta and the proton-neutron mass difference.  
We find that the behavior of the mixing amplitude is very 
complicated in the spacelike region and quite sensitive to 
the proton fraction of nuclear medium and nuclear density. 
In particular, in neutron rich
nuclei (like Pb) and/or neutron stars the mixing amplitudes become
about 10 $\sim$ 100 times as large as those in typical mirror
nuclei. 

\noindent (Keywords: Charge symmetry breaking, Meson miximg,
Asymmetric nuclear matter, Quantum Hadrodynamics)
\end{abstract}

\vspace{1cm}

Charge independence and charge symmetry (CS) provide powerful tools
in organizing and describing the multiplet structure of hadrons and
nuclei.  These symmetries are imperfect and, in particular, charge
symmetry breaking (CSB) has been studied for a long time~\cite{review}.
It is trivially broken by the electromagnetic interaction.  Other well
known sources are the proton($p$)-neutron($n$) mass difference and the
nucleon-nucleon (NN) potential caused by meson mixing, and these
sources can be 
related to the $u$-$d$ quark mass difference in QCD.  Thus, the small
imperfection of CS provide a unique opportunity to investigate the
relation between QCD and hadronic and nuclear observables. 

It has long been recognized that the $\rho$-$\omega$ mixing amplitude
would give rise to the significant part of CSB in the NN force.  Indeed,
substantial effects have been observed in the $e^+ e^- \to \pi^+ \pi^-$
reaction at $q^2 \sim m_\omega^2$ ($m_\omega=782.57$ MeV,
the $\omega$ meson mass)~\cite{review,roexp}. These results allow an
extraction of the
$\rho$-$\omega$ mixing amplitude, $\langle \rho | H_{mix} | \omega
\rangle \simeq - 4500$ MeV$^2$~\cite{roexp}, which seems to be consistent
with CSB in the NN scattering lengths~\cite{review,roexp}, the
Okamoto-Nolen-Schiffer (ONS) anomaly~\cite{review,ons} and the
analyzing powers measured in elastic ${\vec n}$-${\vec p}$ 
scattering~\cite{triumf,iucf}.

However, about a decade ago, the importance of the momentum
dependence of the $\rho$-$\omega$ mixing amplitude was pointed
out using several different models~\cite{off,piwil}.  It suggests that the
$\rho$-$\omega$ mixing amplitude at spacelike momenta is different
from its value at the $\omega$ pole.  A QCD sum-rule calculation also
gives a strong momentum dependence of the amplitude~\cite{qcd}.  It
implies that the NN potential given by the off-shell amplitude of the 
$\rho$-$\omega$ mixing is
quite different from the ones used in the successful phenomenology.

On the other hand, another analysis~\cite{review} argues that the strong 
off-shell effect is inconsistent with the observed $q^2$-dependence of
$\rho$-$\gamma^*$ coupling (see also Ref.~\cite{coon}).  
Moreover, Cohen and Miller~\cite{comil} have
emphasized that simple knowledge of the off-shell meson propagator is
not sufficient to determine the CSB NN potential and that it is important
to know the vertex function computed from the same theory that
supplied the propagator at the same time.  They have showed that
the momentum dependence of the $\rho$-$\omega$ mixing amplitude 
may be absorved into a re-definition of the $\rho$-N vertex.
However, this problem is still controversial~\cite{problem}. 

In this Letter, our aim is to study the properties of CSB meson
mixing in isospin asymmetric nuclear matter.  
To understand CSB in nuclei (for example, the ONS anomaly, nuclear 
structure of heavy nuclei like Pb, neutron stars etc.), 
it is vital to know how CSB effects depend on nuclear density
($\rho_B$) and numbers of proton ($Z$) and neutron ($N$).  Such a
study has not yet been done in the spacelike region~\cite{time}.  
In the present calculation, we use a purely hadronic model, i.e., 
Quantum Hadrodynamics (QHD), to evaluate various meson mixing
amplitudes in asymmetric nuclear matter.  

According to Ref.~\cite{comil}, let us introduce a typical scale 
$\Lambda_s$ 
which separates the short ranged effects from the long ranged ones in
meson exchange models.  Then, the short ranged parts are handled by 
vertex functions, while the long ranged ones are explicitly treated by
meson propagators.  
Here $\Lambda_s$ is taken to be of order of 1 GeV so that
$\sigma$, $\rho$, $\omega$ and $\delta$ (or $a_0$) mesons 
are explicitly included 
but heavier mesons are not. (In the present calculation, the pseudoscalar
mesons are not included because we suppose the relativistic
Hartree approximation (RHA).) 

The Lagrangian density is thus written as
\begin{equation}
{\cal L} = {\cal L}_{QHD-I} + {\cal L}_{NN\delta} + {\cal L}_{NN\rho}
+ \delta{\cal L}_{CT} ,  \label{lag}
\end{equation}
where ${\cal L}_{QHD-I}$ is the usual QHD-I Lagrangian~\cite{serot} 
which describes the $\sigma$-$\omega$-N system.  
The last term in the right hand side 
is a counterterm for renormalizations, which will be discussed below.

The $\delta$ meson is expressed by the Lagrangian density
${\cal L}_{NN\delta}$ and it interacts with the nucleon through
an isovector, scalar coupling 
\begin{equation}
{\cal L}_{NN\delta}^{int.} = g_{\delta} {\bar \psi} \tau_z \delta \psi ,
\label{delint}
\end{equation}
where $\delta$ is the neutral $\delta$-meson field and $g_\delta$ is the
$\delta$-N coupling constant.
The nucleon mass in matter $M_i^*$ ($i=p$ or $n$), which
is a function of the $\sigma$ and $\delta$ fields, is now given by
\begin{equation}
M_{p \choose n}^* = M_{p \choose n} - g_\sigma \sigma \mp g_\delta
\delta ,
\label{pnmass}
\end{equation}
with $\sigma$ the $\sigma$ field, $g_\sigma$ its coupling constant 
and $M_i$ the free proton ($M_p=938.27$MeV) or neutron ($M_n=
939.57$MeV) mass. 

The isovector $\rho$ meson has a vector as well as a tensor coupling
to the nucleon.  The $\rho$-N interaction, which is involved in
the Lagrangian density ${\cal L}_{NN\rho}$, is given by
\begin{equation}
{\cal L}_{NN\rho}^{int.} = g_{\rho} {\bar \psi} \gamma^{\mu}
\rho_{\mu} \tau_z \psi + i \frac{f_{\rho}}{2M_i} {\bar \psi} \sigma^{\mu\nu}
\partial_{\nu} \rho_{\mu}\tau_z \psi ,  \label{nrint}
\end{equation}
where $\psi$ and $\rho^{\, \mu}$ are, respectively, the fields of the
nucleon and the neutral member of $\rho$ meson.  Here 
$g_\rho$ and $f_\rho$ are the vector and tensor
coupling constants, respectively.

To study the meson mixing amplitude, we first have to solve the 
nuclear ground state within RHA. In case of the $\rho$ and $\omega$ mesons, 
since the vector self-energy due to the nucleon loop vanishes in
RHA~\cite{serot,saito}, the vacuum polarization does not change the energy 
density of the total system. 

{}For the scalar mesons, it is necessary to 
treat the $\sigma$-$\delta$ mixing and the polarization corrections to 
the $\sigma$- and $\delta$-meson propagators simultaneously.  
In Eq.(\ref{lag}), $\delta{\cal L}_{CT}$ involves counterterms
such as $\sigma^n \delta^m$ ($n,m = 1 \sim 4$), which remove the divergent
pieces of the nucleon-loop contributions to the $\sigma$-$\delta$
mixing and the scalar-meson propagators.  
As in Ref.~\cite{serot}, the counterterms are determined so as to
renormalize the polarization functions at $q_\mu^2 = 0$ and $M_i^*=M_i$.
This prescription provides a finite vacuum
fluctuation correction to the energy density of the total system.  We
finally find that in RHA the correction is given by a sum of 
$\Delta {\cal E}_p$ and $\Delta {\cal E}_n$, where
\begin{eqnarray}
\Delta{\cal E}_i &=& -\frac{1}{8\pi^2} \left[ M_i^{*4} \ln \left(
\frac{M_i^*}{M_i} \right) + M_i^3(M_i-M_i^*) - \frac{7}{2}M_i^2
(M_i-M_i^*)^2 \right. \nonumber \\
&+& \left. \frac{13}{3}M_i(M_i-M_i^*)^3 - \frac{25}{12}
(M_i-M_i^*)^4 \right] .
\label{delE}
\end{eqnarray}

The total energy density in RHA is thus given by
\begin{equation}
{\cal E}_{tot} = {\cal E}_{MFA} + \Delta{\cal E}_p + \Delta{\cal E}_n ,
\label{totalE}
\end{equation}
where ${\cal E}_{MFA}$ is the energy density in the mean-field
approximation (MFA)~\cite{serot}.  
In the present calculation, the $\sigma$, $\delta$
and $\rho$ masses are, respectively, taken to be
$m_\sigma = 550.0$ MeV, $m_\delta = 983.0$ MeV and
$m_\rho = 769.3$ MeV.  We determine $g_\sigma$ and $g_\omega$
(the $\omega$-N coupling constant) so as to fit the saturation condition
for symmetric ($Z=N$) nuclear matter: ${\cal E}_{tot}/\rho_B - M = -15.75$
MeV ($M$, the average of the free proton and neutron masses) at the
saturation density $\rho_0=0.17$ fm$^{-3}$.  Fixing the ratio of the
tensor to vector coupling constants, $c_\rho \equiv f_\rho / g_\rho$, to
be 5.0~\cite{bonn}, the vector coupling
constant $g_\rho$ is chosen so as to reproduce the measured
$\rho$-$\omega$ mixing amplitude, 
$\langle \rho |H_{em}|\omega \rangle = -4520$ MeV$^2$, at the 
$\omega$ pole~\cite{roexp}.  Finally, we
determine $g_\delta$ so as to produce the empirical symmetry energy 
$a_4 (= 32.5$ MeV)~\cite{moz}.
We then find that $g_\sigma =8.360$,
$g_\delta =3.257$, $g_\omega =9.493$ and $g_\rho =3.455$.
This yields the effective proton and neutron masses, $M_p^* =679.36$ MeV
and $M_n^* =680.65$ MeV, at $\rho_0$ and the nuclear incompressibility
$K = 460$ MeV.

Now we are in a position to calculate meson mixing amplitudes 
in asymmetric ($Z \neq N$) nuclear matter.  There are four types of
the CSB meson mixing: $\rho$-$\omega$, $\sigma$-$\delta$, $\sigma$-$\rho$
and $\delta$-$\omega$ mixing.  In particular, the mixing of scalar and
vector mesons (the latter two cases) is similar to the familiar 
$\sigma$-$\omega$ mixing in matter~\cite{saito,chin}.  It is a purely
density effect and forbidden in vacuum.  Note that the
$\rho$-$\omega$ mixing yields class III and IV forces
while the others provide class III forces~\cite{review}. 

We first calculate the $\rho$-$\omega$ mixing.  The polarization
function due to the nucleon loop has a tensor character
and it can be separated into the longitudinal (L) and transverse (T)
components.  They are identical each other in vacumm.  The 
mixing amplitudes for the L and T modes are, respectively, given by
the polarization functions as 
\begin{equation}
\langle \rho | H_{mix} | \omega \rangle_{L} = \frac{q_\mu^2}{q_s^2}
\Pi^{00}_{\rho \omega}(q_0,q_s) \ \ \ \mbox{and} \ \ \
\langle \rho | H_{mix} | \omega \rangle_{T} = \frac{1}{2}
[\Pi^{11}_{\rho \omega}(q_0,q_s) + \Pi^{22}_{\rho \omega}(q_0,q_s)] ,
\label{roLT}
\end{equation}
where we choose the direction of the meson momentum as 
$q^\mu=(q_0,0,0,q_s)$ and
\begin{equation}
\Pi^{\mu\nu}_{\rho \omega}(q_0,q_s) =
-i g_\omega g_\rho \int\frac{d^4k}{(2\pi)^4}\mbox{Tr}[\gamma^{\mu}
G(k+q) \tilde{\Gamma}^{\nu}_\rho \tau_z G(k)] .
\label{ropol}
\end{equation}
Here $G$ is the nucleon propagator~\cite{serot} and is a diagonal
matrix in isospin space
\begin{equation}
G(k)=\left(\begin{array}{cc}
 G_{p}(k) & 0
\\ 0 & G_{n}(k)
\end{array}
\right) .
\label{Nprop}
\end{equation}
As usual, the nucleon propagator can be devided into the Feynman (F)
and density-dependent (D) parts~\cite{serot}.

In the $\rho$-N tensor coupling in matter, it is natural to 
replace the free nucleon mass in 
Eq.(\ref{nrint}) with the effective mass $M_i^*$ 
because it is originally derived by the Gordon 
decomposition of the nucleon field in matter\footnote{
This replacement 
of the nucleon mass is equivalent to adding terms such as
$\sigma^n {\bar \psi} \sigma^{\mu\nu} \partial_{\nu}
\rho_{\mu} \tau_z \psi$ and $\delta^n {\bar \psi} \sigma^{\mu\nu}
\partial_{\nu} \rho_{\mu} \tau_z \psi$ ($n = 1, 2, \cdots, \infty$) to 
the original Lagrangian density.  However, those terms do not modify 
the $\sigma$ and $\delta$ fields in RHA.}.
Hence, in Eq.(\ref{ropol}) we use the $\rho$-N coupling as
\begin{equation}
\Gamma^{\mu}_\rho = \gamma^{\mu}
- i \frac{c_\rho}{2M_i^*} \sigma^{\mu\lambda}q_\lambda 
\ \ \ \mbox{and} \ \ \ 
\tilde{\Gamma}^{\mu}_\rho = \gamma^{\mu}
+ i \frac{c_\rho}{2M_i^*} \sigma^{\mu\lambda}q_\lambda . 
\label{Gam}
\end{equation}

Because of the isospin matrix $\tau_z$, 
the mixing amplitude is given by the difference between contributions
from the proton and neutron loops. The polarization function which
involves only the F part of $G$ becomes divergent.  However, as in 
Ref.~\cite{piwil}, by virtue of the replacement in Eq.(\ref{Gam}) the
difference between proton and neutron contributions becomes 
finite even in matter.  The F part of the polarization function 
is then given by a sum of
$\Pi^{F\mu\nu}_{v}$ and $\Pi^{F\mu\nu}_{t}$, where the former is the
polarization function with the vector coupling at the $\rho$-N vertex
while the latter is that with the tensor coupling: 
\begin{eqnarray}
\Pi^{F\mu\nu}_{v}(q_\mu^2)
&=& \frac{\xi_{\mu\nu}}{2\pi^2}g_\omega g_\rho q_\mu^2 \int_0^1 dx x(1-x) \ln
\left[ \frac{M_p^{*2}-x(1-x)q_\mu^2}{M_n^{*2}-x(1-x)q_\mu^2} \right] ,
\label{ropolvv} \\
\Pi^{F\mu\nu}_{t}(q_\mu^2)
&=& \frac{\xi_{\mu\nu}}{8\pi^2} g_\omega f_\rho q_\mu^2 \int_0^1 dx \ln
\left[ \frac{M_p^{*2}-x(1-x)q_\mu^2}{M_n^{*2}-x(1-x)q_\mu^2} \right] ,
\label{ropolvt}
\end{eqnarray}
with $\xi^{\mu\nu} = -g^{\mu\nu} + q^\mu q^\nu /q_\mu^2$.  The
polarization function involving at least one power of the D part
of $G$ can be calculated analytically~\cite{saito}.

In the present calculation, we introduce a monopole-type form
factor at each vertex in the spacelike region.  Thus, we replace the 
coupling constant as
\begin{equation}
g \to g(q_\mu^2) = g / (1 - q_\mu^2 / \Lambda^2 ) ,
\label{ff}
\end{equation}
with the cutoff $\Lambda = 1.5$ GeV.  Here we suppose that $\Lambda$ 
is common to all the vertices.  (As pointed out in
Ref.~\cite{comil}, properly speaking, it is necessary to calculate the
CSB vertex from the same QHD theory that provided the propagator.) 

\begin{figure}
\begin{center}
\epsfig{file=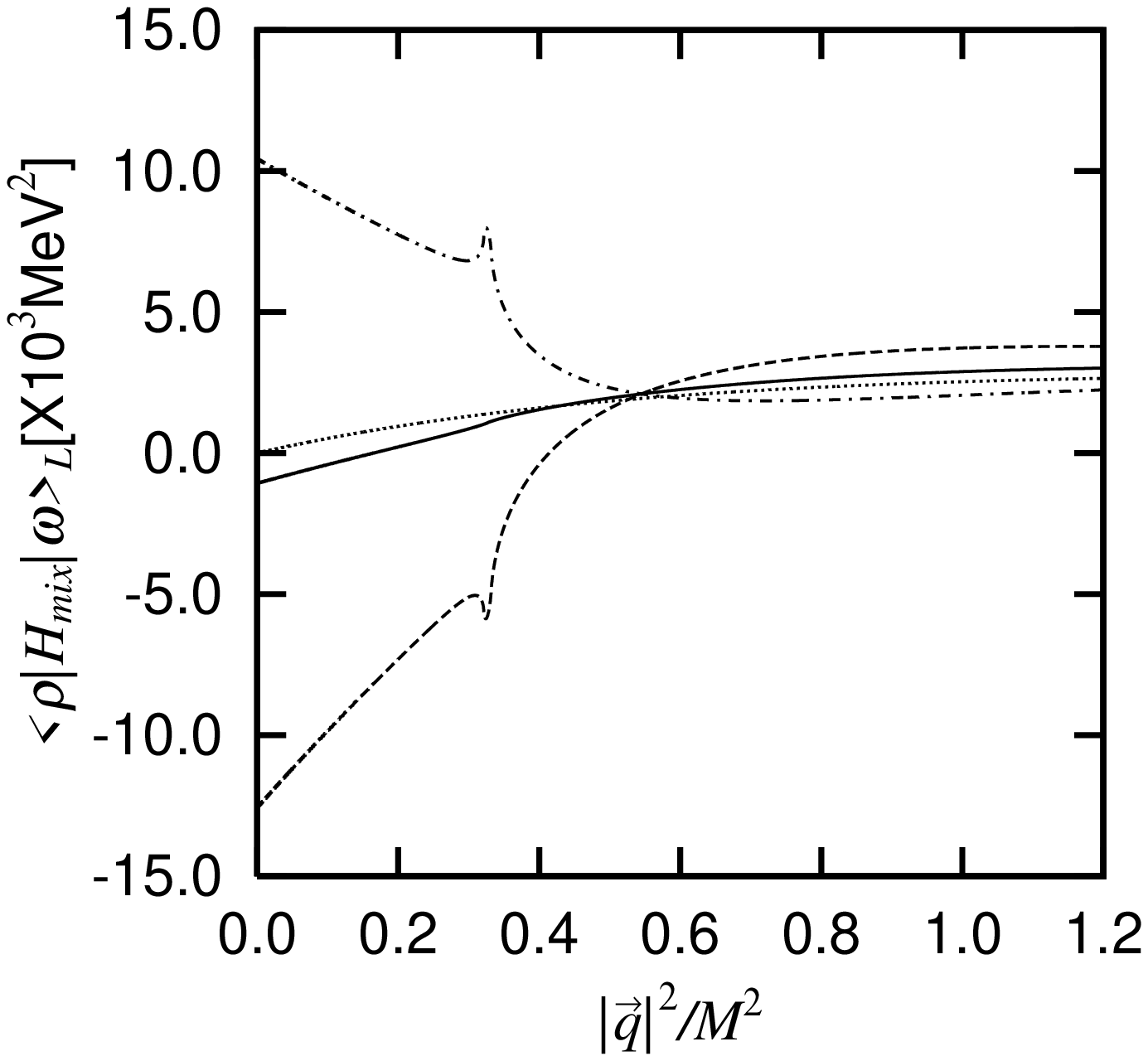,height=10cm}
\epsfig{file=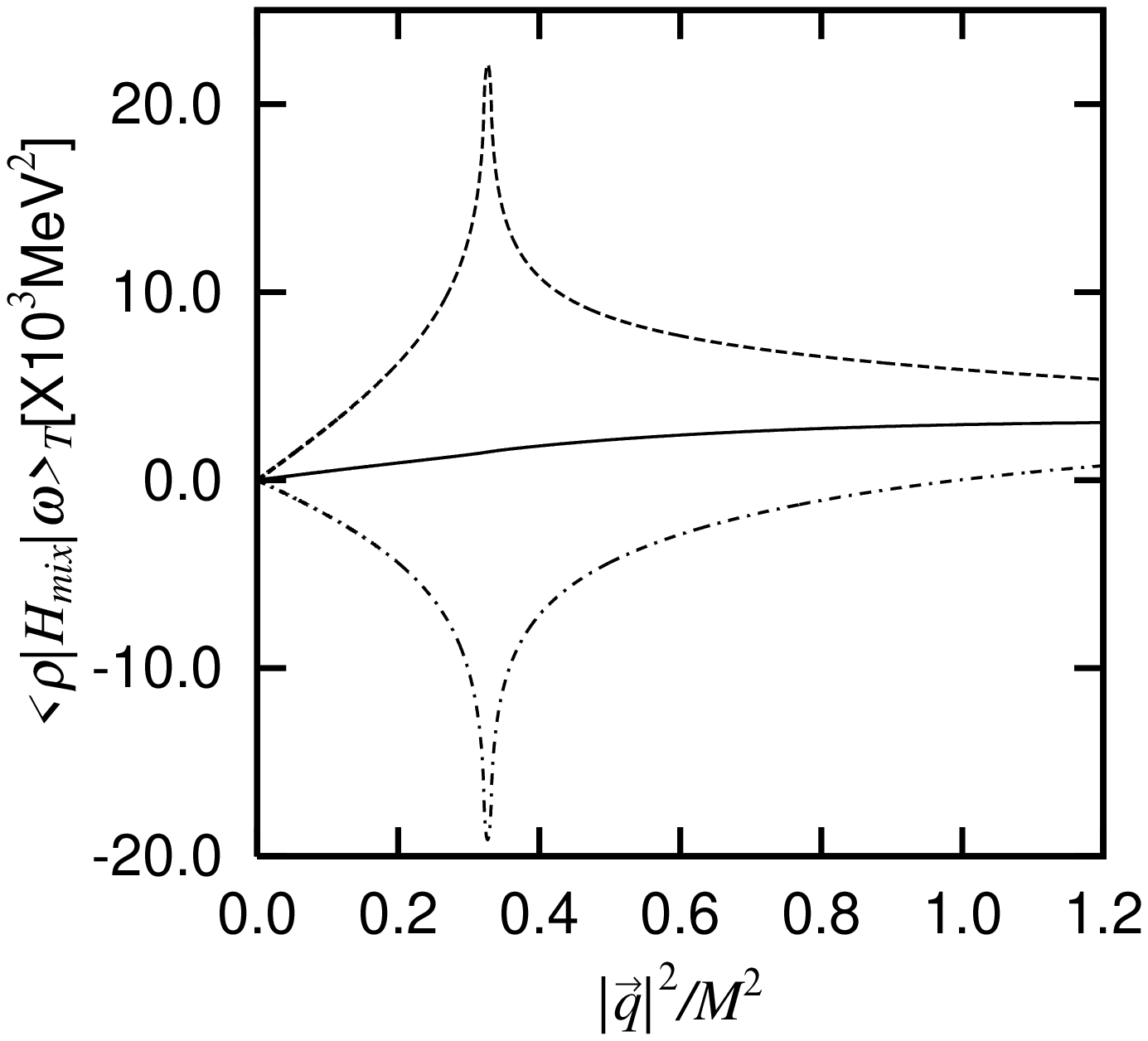,height=10cm}
\caption{
Amplitude of the $\rho$-$\omega$ mixing ($\rho_B = \rho_0$) 
in the spacelike region. We take $q_0 = 0$ and vary $|{\vec q}| (=q_s)$.  
The top panel is for the L mode while the bottom one is for the 
T mode. The solid (dashed) [dot-dashed] curve represents the result 
with $f_p = 0.5 (0.49) [0.51]$ while the result in vacuum is shown 
by the dotted curve in the top panel.  
}
\label{f:rolt}
\end{center}
\end{figure}

\begin{figure}
\begin{center}
\epsfig{file=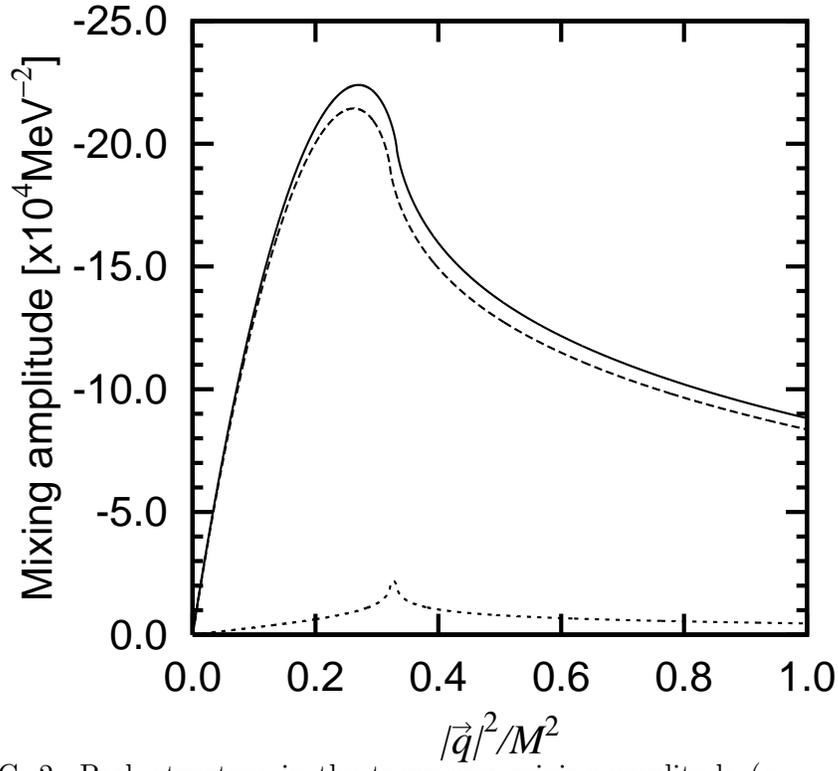,height=10.5cm}
\caption{
Peak structure in the transverse mixing amplitude ($\rho_B = \rho_0$ 
and $f_p = 0.51$). 
The solid (dashed) curve represents the mixing amplitude due to the  
proton (neutron) loop while the difference between proton and neutron
contributions are shown by the short-dashed curve.  
}
\label{f:peak}
\end{center}
\end{figure}

\begin{figure}
\begin{center}
\epsfig{file=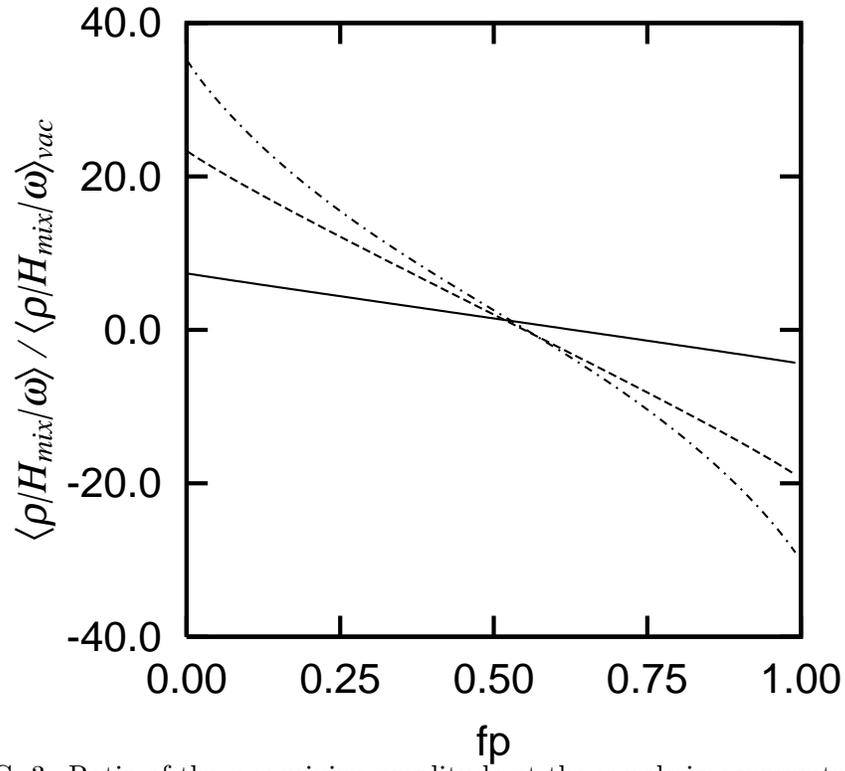,height=10.5cm}
\caption{
Ratio of the $\rho$-$\omega$ mixing amplitude at the $\omega$ pole in
asymmetric nuclear matter to
that in vacuum.  The solid (dashed) [dot-dashed] curve is for $\rho_0$
(2$\rho_0$) [3$\rho_0$].  
}
\label{f:onshell}
\end{center}
\end{figure}

In Fig.~\ref{f:rolt}, we present the $\rho$-$\omega$ mixing
amplitudes for the L and T modes in the spacelike region. 
We fix $q_0=0$ and vary $|{\vec q}| (=q_s)$.  
The nuclear density is chosen to be $\rho_0$. 
Here we introduce the proton fraction $f_p$ that is defined by $Z/(Z+N)$.  
In the figure, in addition to the case of symmetric nuclear matter 
($f_p = 0.5$), the results for $f_p = 0.49$ and $0.51$, which 
respectively correspond to 
the mirror nuclei of $^{41}$Ca and $^{41}$Sc, are illustrated.  
The mixing amplitude in symmetric matter is very similar to that in 
vacuum.\footnote{
The present result in vacuum is consistent with the result calculated 
in Ref.~\cite{piwil}.}
However, in asymmetric matter the amplitude is much larger
than that in symmetric case.  The 
mixing amplitude for $f_p = 0.51$ is opposite in sign relative to that
for $f_p = 0.49$ because the amplitude is given by the difference
between proton and  neutron contributions.  It should be noticed 
that there is a sharp peak around $q_s = 2 k_F^i$ ($k_F^i$, the Fermi
momentum of proton or neutron) in the mixing amplitude and that the peak
structure stems from the D part of the polarization function.  As an
example, in Fig.~\ref{f:peak} we illustrate how such a peak is
generated in the T mode with $f_p = 0.51$.  From
the figure, we can see that the peak is produced by the cancellation of
large amplitudes from the proton and neutron loops.  
In Fig.~\ref{f:onshell}, as a function of $\rho_B$ and $f_p$, 
we show the ratio of the $\rho$-$\omega$
mixing amplitude at the $\omega$ pole ($q_0 = m_\omega$ and $q_s = 0$)
in asymmetric matter to that in vacuum.  Note that in 
the limit $q_s \to 0$ ($q_0 \neq 0$) the L mode is identical to 
the T mode. The mixing amplitude at $\rho_0$ decreases gradually as $f_p$ 
increases. However, at high $\rho_B$ the amplitude changes
drastically. In particular, at small $f_p$ and high $\rho_B$ the ratio
becomes $30 \sim 40$.  Such an extreme condition may correspond to the
case of neutron stars~\cite{moz,nstar}. 

Next we consider the $\sigma$-$\delta$ mixing.  
The mixing amplitude (or the polarization function) is given by
\begin{equation}
\langle \delta | H_{mix} | \sigma \rangle = \Pi_{\sigma \delta}(q_0,q_s)
= -i g_{\sigma} g_{\delta} \int\frac{d^4k}{(2\pi)^4}\mbox{Tr}
[G(k+q)\tau_z G(k)] . 
\label{sigdel}
\end{equation}
The amplitude is again given by the difference between contributions
from the proton and neutron loops.  As mentioned above Eq.(\ref{delE}),
the $\sigma$-$\delta$ mixing does not vanish in vacuum if $M_p \neq M_n$.
The normalized F part of the polarization function 
is given by $\Pi^F_{\sigma\delta}(q^2_\mu) = 
\Pi^F_{\sigma\delta p}(q^2_\mu) - \Pi^F_{\sigma\delta n}(q^2_\mu)$, 
where~\cite{fh} 
\begin{eqnarray}
\Pi^F_{\sigma\delta i}(q^2_\mu) &=& \frac{3g_{\sigma}g_{\delta}}{4\pi^2} 
\left( M^2_i
+ 3M^{*2}_i-4M^*_iM_i-\frac{1}{6}q^2_\mu \right. \nonumber \\
&-& \int_{0}^{1}dx \left. [M^{*2}_i-x(1-x)q^2_\mu]\ln \left[ 
\frac{M^{*2}_i-x(1-x)q^2_\mu}{M^2_i} \right] \right) . 
\label{polsd}
\end{eqnarray}
The D part of the polarization function can be calculated explicitly. 
It is similar to the D part of the nucleon-loop
contribution to the $\sigma$-meson propagator, except for the isospin
matrix $\tau_z$~\cite{saito}.

{}For the scalar-vector ($\sigma$-$\rho$ or $\delta$-$\omega$) meson
mixing, the mixing amplitude (or the polarization function) is given by
\begin{equation}
\langle v | H_{mix} | s \rangle = \Pi_{sv}(q_0,q_s)
= i g_s g_v \delta_{0\mu} \int \frac{d^4k}{(2\pi)^4}
\mbox{Tr} [G(k+q)\tilde{\Gamma}^{\mu}_v\tau_z G(k)] ,
\label{sv}
\end{equation}
where $s$ and $v$ stand for the scalar and vector mesons, respectively 
(if $v=\omega$, $\tilde{\Gamma}^{\mu}_{\omega} = \gamma^\mu$). 
As in the $\sigma$-$\omega$ mixing~\cite{saito,chin}, this amplitude
vanishes identically in vacuum and hence is a purely density effect.  
The polarization function can be calculated analytically~\cite{saito}.

\begin{figure}
\begin{center}
\epsfig{file=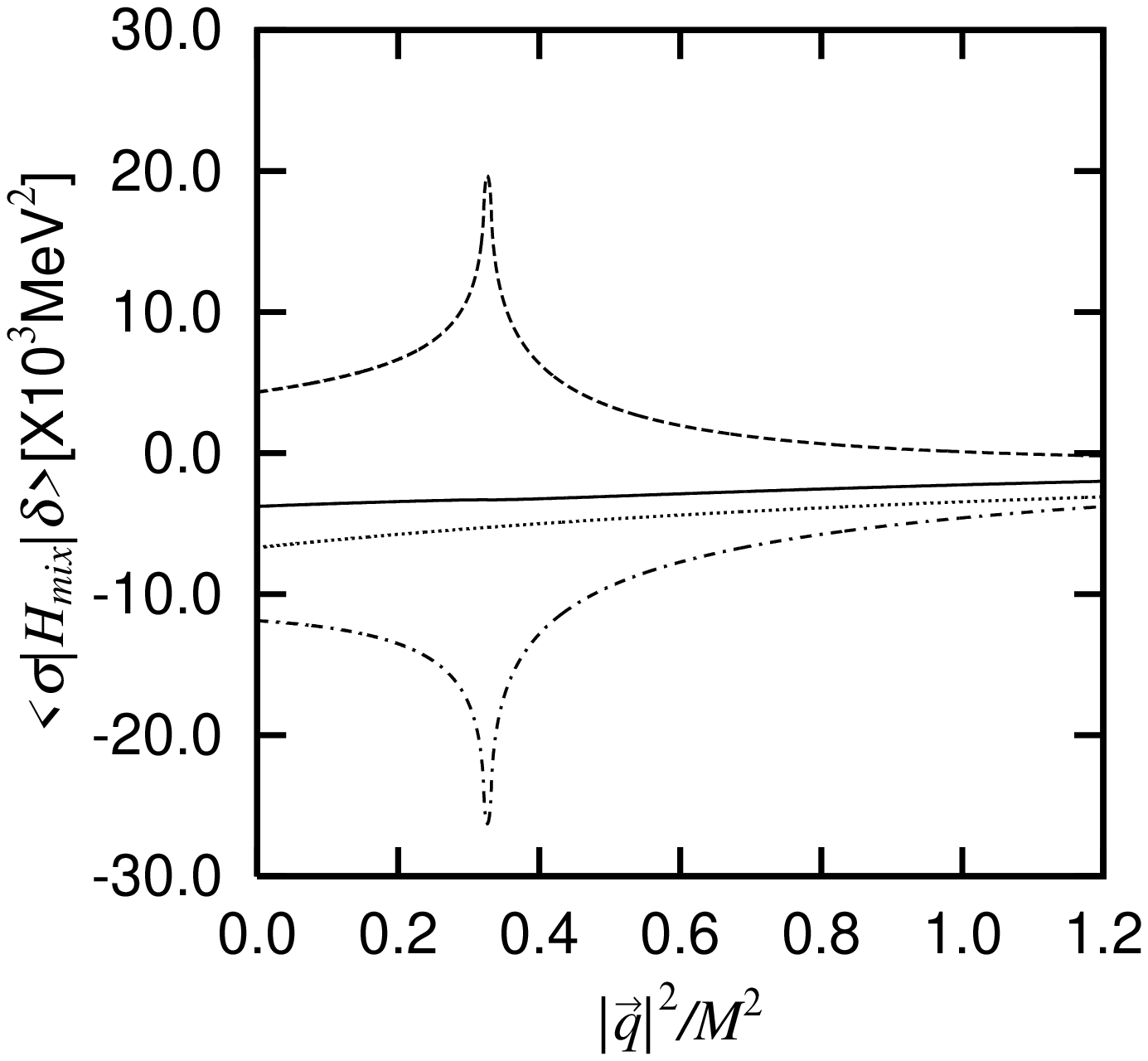,height=7.3cm}
\epsfig{file=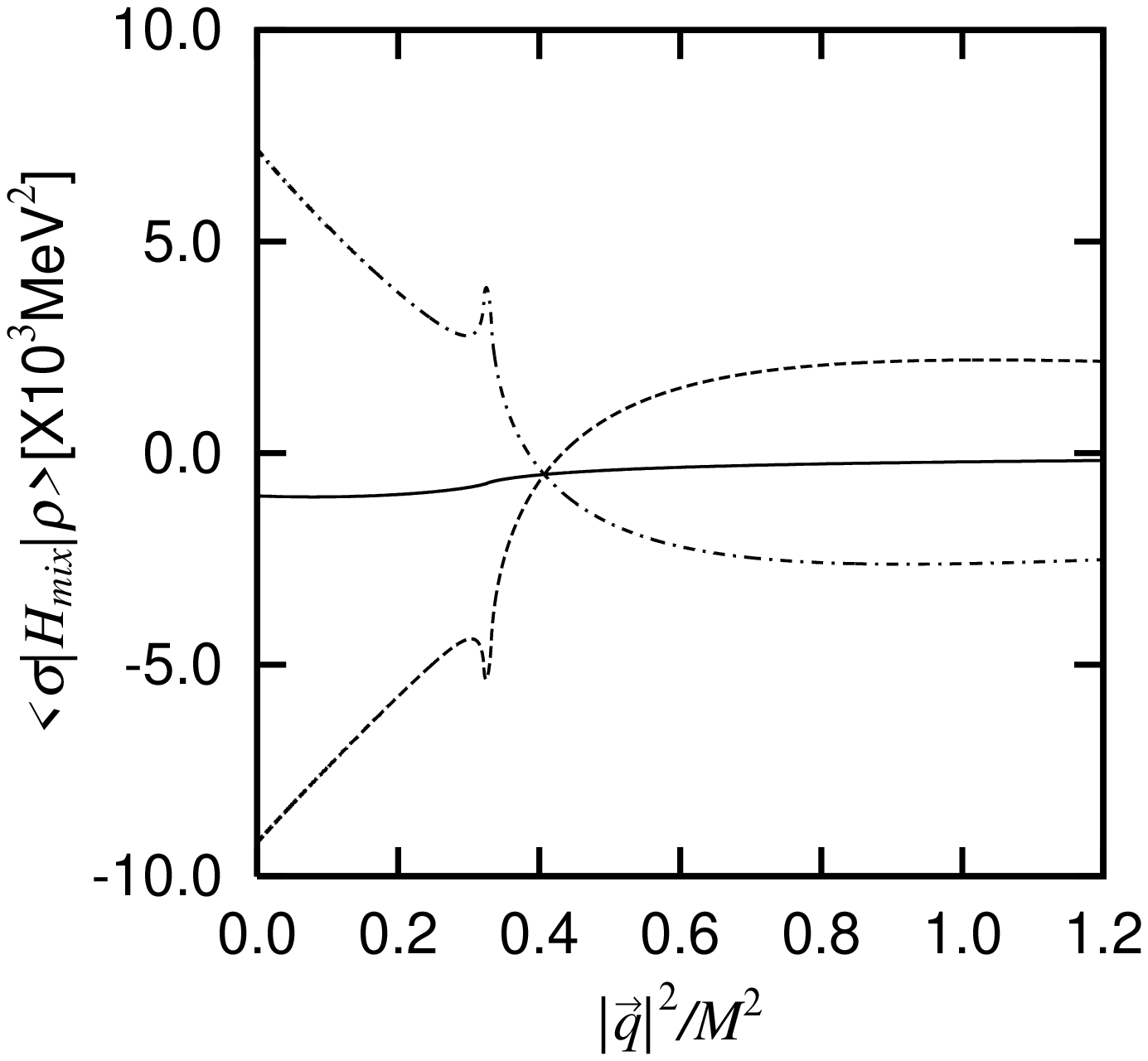,height=7.3cm}
\epsfig{file=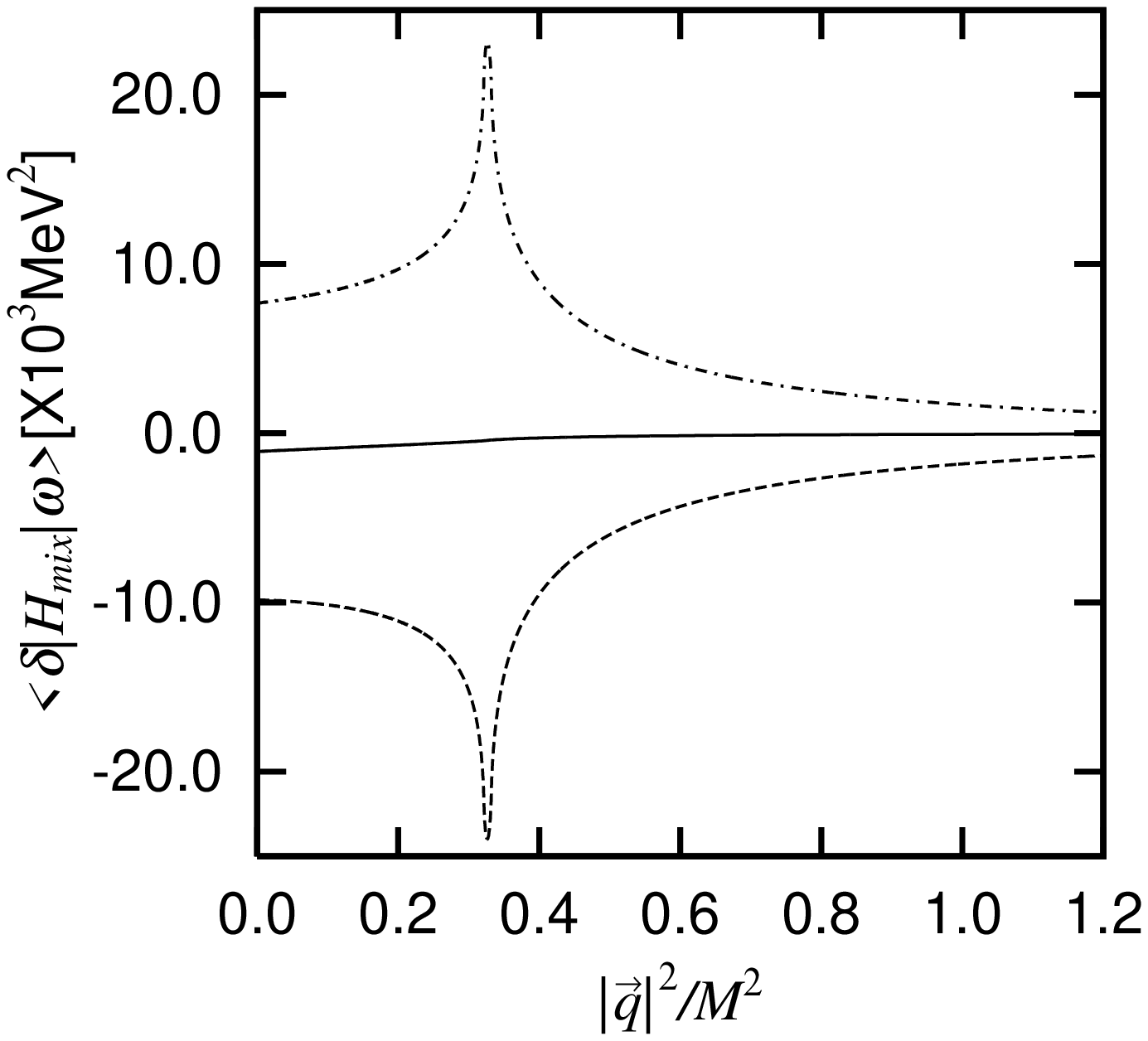,height=7.3cm}
\caption{
Same as Fig.~\protect\ref{f:rolt} but for the $\sigma$-$\delta$ (top panel), 
$\sigma$-$\rho$ (middle panel) and $\delta$-$\omega$ (bottom panel)
mixing amplitudes ($\rho_B = \rho_0$).  
}
\label{f:other}
\end{center}
\end{figure}

In Fig.~\ref{f:other}, we present the results of the $\sigma$-$\delta$,
$\sigma$-$\rho$ and $\delta$-$\omega$ mixing in asymmetric nuclear
matter. The behavior of the $\sigma$-$\delta$ mixing is very 
similar to that of the $\delta$-$\omega$ mixing.  However, the  
$\sigma$-$\delta$ mixing amplitude is opposite in sign compared with  
the $\delta$-$\omega$ mixing. 
By contrast, in the $\sigma$-$\rho$ mixing 
the sign of the amplitude changes at $q_s \simeq 0.63 M$.  This is 
due to the $\rho$-N tensor coupling that dominates for large 
$q_s$. In the top panel of the figure, the $\sigma$-$\delta$ mixing
amplitude in vacuum is also shown by the dotted curve.  It may also 
contribute to the NN scattering lengths in free space~\cite{review}.  

\begin{figure}
\begin{center}
\epsfig{file=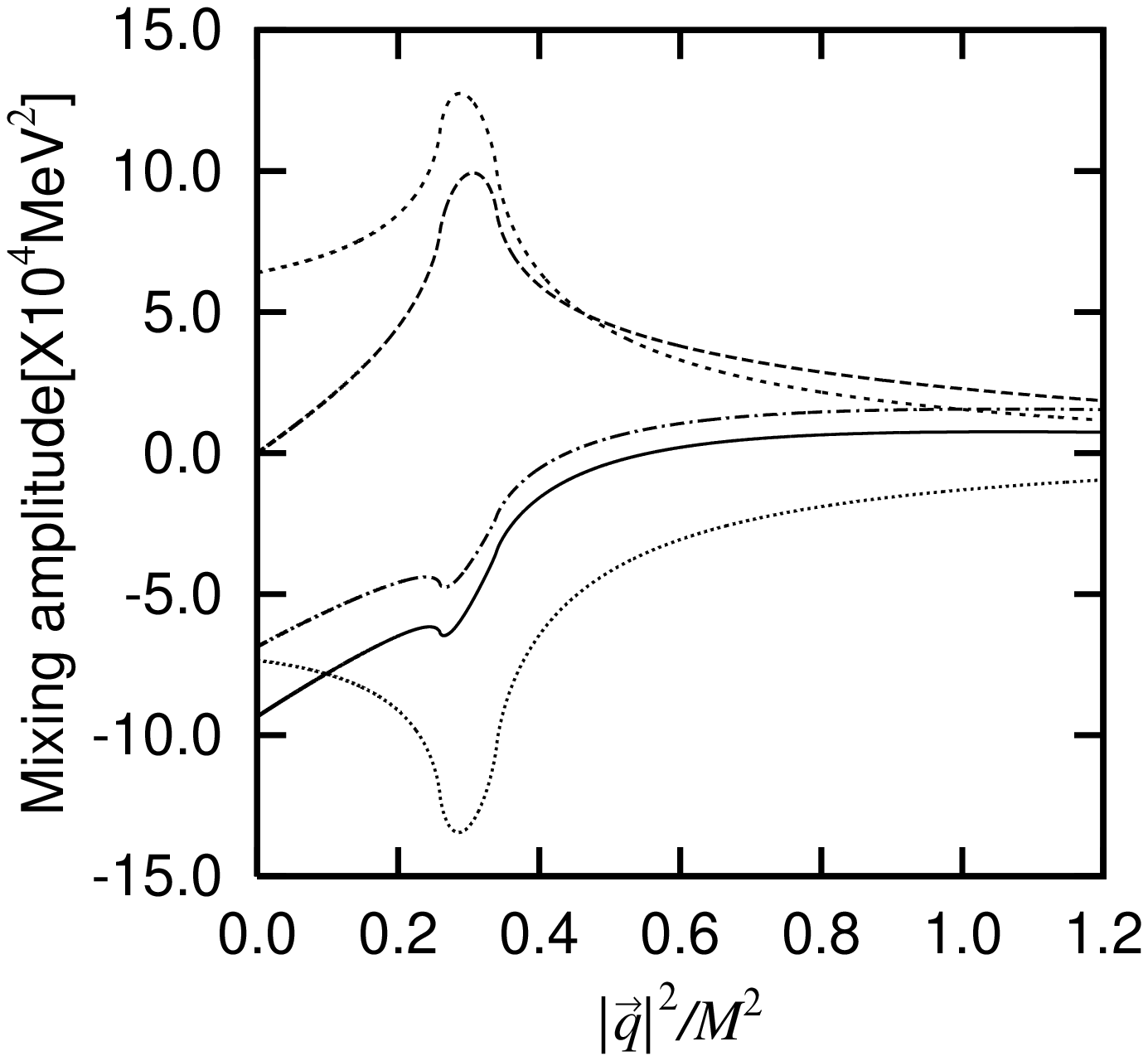,height=10cm}
\epsfig{file=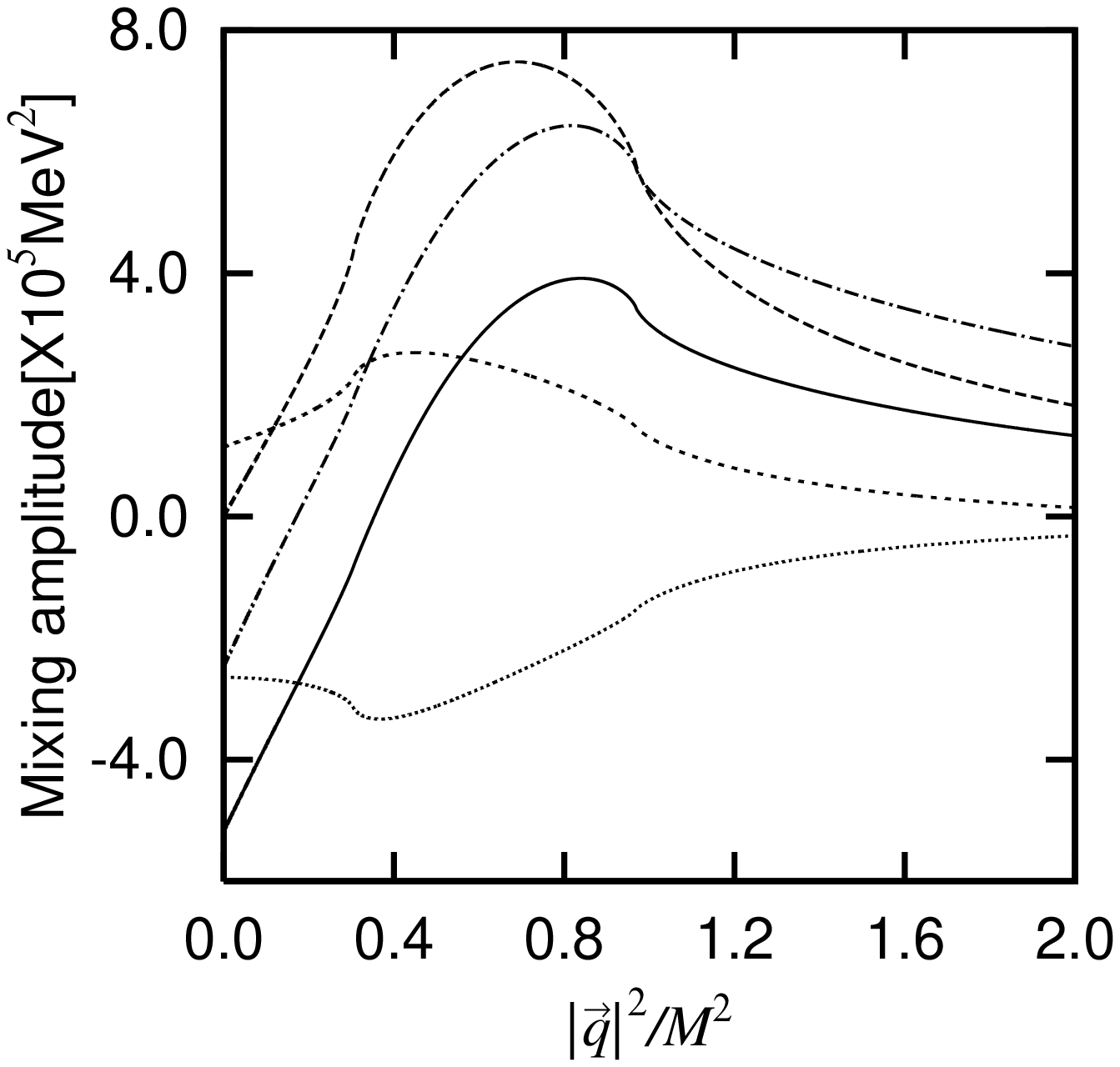,height=10cm}
\caption{
Meson mixing amplitudes in Pb (top panel) and neutron stars (bottom panel). 
We take $\rho_B = 0.15$ fm$^{-3}$ and $f_p = 0.4$ for Pb while 
$\rho_B = 3 \rho_0$ and $f_p = 0.15$ for neutron stars.
The L (T) mode of the $\rho$-$\omega$ mixing is shown by the solid
(long-dashed) curve.  The short-dashed (dotted) [dot-dashed] curve is
for the $\sigma$-$\delta$ ($\delta$-$\omega$) [$\sigma$-$\rho$]
mixing. 
}
\label{f:pbstar}
\end{center}
\end{figure}

In Fig.~\ref{f:pbstar}, we illustrate the meson mixing amplitues in Pb 
($\rho_B = 0.15$ fm$^{-3}$ and $f_p = 0.4$) and neutron stars. 
We suppose that in neutron stars the nuclear density is about 
$3 \rho_0$ and $f_p \sim 0.15$~\cite{moz,nstar}.  From the figure we
can see that in Pb the mixing amplitudes are  
about $5 \sim 10$ times as large as those in matter with $f_p = 0.49$ 
(see Figs.~\ref{f:rolt} and \ref{f:other}).  Furthermore, in neutron
stars the $\rho$-$\omega$ mixing amplitude in the L mode and the
$\sigma$-$\rho$ mixing amplitude are about 100 times larger than
those for $f_p = 0.49$.  
This fact is consistent with the result shown in
Fig.~\ref{f:onshell}.  

In summary, using QHD we have studied the $\rho$-$\omega$,
$\sigma$-$\delta$, $\sigma$-$\rho$ and $\delta$-$\omega$ mixing
amplitudes in isospin asymmetric nuclear matter.  We have
shown that the mixing amplitude is quite sensitive to the proton  
fraction and nuclear density. 
In particular, in neutron stars and/or neutron rich
nuclei (like Pb) the CSB meson mixing amplitudes become considerably 
large compared with those in typical mirror nuclei. 
Even in light nuclei, if $f_p$ is far from 0.5 (like halo
nuclei~\cite{halo}), the CSB meson mixing would play an important role 
in the nuclear structure.
In the present calculation, we have not calculated the CSB potential 
generated by the meson mixing because, in addition to the mixing
amplitude, it is necessary to know the CSB meson-N vertex to
draw any definite conclusions on the potential~\cite{comil}.  Such a 
complete study requires more elaborate calculations and it is beyond
the scope of this Letter.  It would clearly be very interesting and
important to calculate it in the future. 

\vspace{0.5cm}
We would like to thank A. Suzuki, A.W. Thomas and A.G. Williams for
valuable discussions.




\newpage

\newpage

\newpage
\newpage

\newpage


\newpage
\begin{thebibliography}{99}
%
\bibitem{review} E.M. Henley, G.A. Miller, in: {\it Mesons in nuclei},   
edited by M. Rho and 
D.H. Wilkinson (North-Holland, Amsterdam, 1979), p.405; \\
G.A, Miller, B.M.K. Nefkens, I. {\v S}laus, Phys. Rep. 194
(1990) 1; \\
G.A. Miller, W.T.H. Van Oers, in: {\it Symmetries and fundamental
interactions in nuclei}, edited by W.C. Haxton and E.M. Henley (World
Scientific, Singapore, 1995), p.127.
%
\bibitem{roexp} L.M. Barkov {\it et al.}, Nucl. Phys. B256 (1985) 365; \\
S.A. Coon, R.C. Barret, Phys. Rev. C36 (1987) 2189.
%
\bibitem{ons} S. Shlomo, Rep. Prog. Phys. 41 (1978) 957; \\
P.G. Blunden, M.J. Iqbal, Phys. Lett. B198 (1987) 14; \\
K. Saito, A.W. Thomas, Phys. Lett. B335 (1994) 17; \\ 
M. Kimura, A. Suzuki, H. Tezuka, Phys. Lett. B367 (1996) 5; \\
K. Tsushima, K. Saito, A.W. Thomas, Phys. Lett. B465 (1999) 36. 
%
\bibitem{triumf} R. Abegg, Phys. Rev. Lett. 56 (1986) 2571;
  Phys. Rev. D39 (1989) 2464.
%
\bibitem{iucf} L.D. Knutson, Phys. Rev. Lett. 66 (1991) 1410; \\
S.E. Vigdor, Phys. Rev. C46 (1992) 410.
%
\bibitem{off} T. Goldman, J.A. Henderson, A.W. Thomas, Few Body
  Syst. 12 (1992) 193; \\
G. Krein, A.W. Thomas, A.G. Williams, Phys. Lett. B317 (1993) 293 \\
H.B. O'Connell, B.C. Pearce, A.W. Thomas, A.G. Williams, Phys. Lett. 
B336 (1994) 1. 
%
\bibitem{piwil} J. Piekarewicz, A.G. Williams, Phys. Rev. C47 (1993)
  R2462.
%
\bibitem{qcd} T. Hatsuda, E.M. Henley, Th. Meissner, G. Krein,
  Phys. Rev. C49 (1994) 452.
%
\bibitem{coon} S.A. Coon, B.H.J. McKeller, A.A. Rawlinson, in: {\it
Intersections between particle and nuclear physics}, edited by
T.W. Donelly, AIP conf. proc. no.412 (AIP, N.Y., 1997),
p.368.
%
\bibitem{comil} T.D. Cohen, G.A. Miller, Phys. Rev. C52 (1995) 3428. \\
See also, S. Gardner, C.J. Horowitz, J. Piekarewicz, Phys. Rev. C53
(1996) 1143. 
%
\bibitem{problem} H.B. O'Connell, B.C. Pearce, A.W. Thomas,
  A.G. Williams, Prog. Part. Nucl. Phys. 39 (1997) 201. 
%
\bibitem{time} For the timelike region, see, for example,
  A.K. Dutt-Mazumder, Nucl. Phys. A611 (1996) 442.
%
\bibitem{serot} B.D. Serot, J.D. Walecka, Adv. Nucl. Phys. 16 (1986) 1.
%
\bibitem{saito} K. Saito, K. Tsushima, A.W. Thomas, A.G. Williams,
Phys. Lett. B433 (1998) 243; \\
K. Lim, C.J. Horowitz, Nucl. Phys. A501 (1989) 729.
%
\bibitem{bonn} R. Machleidt, in {\it Proceedings of the relativistic
  dynamics and quark-nuclear physics}, edited by M.B. Johnson and
  A. Picklesimer (Wiley, New York, 1986); \\
J.J. Sakurai, in {\it Currents and mesons} (Univ. of Chicago Press,
  Chicago, 1969).  
%
\bibitem{moz} N.K. Glendenning, F. Weber, S.A. Moszkowski, Phys. Rev. C45
(1992) 844.
%
\bibitem{chin} S.A. Chin, Ann. of Phys. (N.Y.) 108 (1977) 301.
%
\bibitem{nstar} F. Weber, astro-ph/0207053; \\
C.J. Horowitz, J. Piekarewicz, nucl-th/0207067.
%
\bibitem{fh} R.J. Furnstahl, C.J. Horowitz, Nucl. Phys. A485 (1988)
  632.
%
\bibitem{halo} I. Tanihata, Nucl. Phys. A654 (1999) 235c. 
%

\end{thebibliography}
\end{document}